# Utilizing Pre-trained Language Models and Large Language Models for 10-K Items Segmentation

**Hsin-Min Lu**
National Taiwan University

**Yu-Tai Chien**
National Taipei University of Business

**Huan-Hsun Yen**
**Yen-Hsiu Chen**
National Taiwan University

**Running Head:** PLMs and LLMs for 10-K Items Segmentation

This work is supported in part by the National Science and Technology Council, Taiwan, under the Grant NSTC 112-2410-H-002-087-MY3, and by the Center for Research in Econometric Theory and Applications under the Grant 113L900203 from The Featured Areas Research Center Program within the framework of the Higher Education Sprout Project by the Ministry of Education (MOE) in Taiwan.

The authors declare no financial conflicts of interest concerning the study presented in this article.

Hsin-Min Lu, National Taiwan University, College of Management, Department of Information Management and the Center for Research in Econometric Theory and Applications, Taipei, Taiwan; Yu-Tai Chien, National Taipei University of Business, College of Management, Department of Business Administration, Taipei, Taiwan; Huan-Hsun Yen and Yen-Hsiu Chen, National Taiwan University, College of Management, Department of Information Management, Taipei, Taiwan.

**Data Availability:** Data available upon request.

Supplemental materials are available.

**Abstract**

Extracting specific items from 10-K reports is challenging due to variations in document formats and item presentation. To improve over traditional rule-based approaches, this study introduces and compares two advanced item segmentation methods: (1) GPT4ItemSeg, using a novel line-ID-based prompting mechanism to utilize a large language model, ChatGPT-4o, for item segmentation, and (2) BERT4ItemSeg, combining a pre-trained language model, BERT, with a Bi-LSTM model in a hierarchical structure to overcome context window constraints. Trained and evaluated on 3,737 annotated 10-K reports, BERT4ItemSeg achieves a macro-F1 of 0.9825, surpassing GPT4ItemSeg (0.9567), conditional random field (0.9818), and rule-based methods (0.9048) for core items (1, 1A, 3, and 7). These approaches enhance item segmentation performance, improving text analytics in accounting and finance. BERT4ItemSeg offers satisfactory item segmentation performance, while GPT4ItemSeg can easily adapt to regulatory changes. Together, they provide an extensible framework for 10-K item segmentation that supports reliable and reproducible results.

## AI Disclosure Statement

During the preparation of this manuscript, the authors used generative artificial intelligence tools, including ChatGPT and Grok, to assist with language refinement, clarification of phrasing, and improvement of overall readability. These tools were used in an advisory capacity only and did not generate original research ideas, data, analyses, results, or interpretations. All AI-generated suggestions were reviewed, edited, and verified by the authors, who take full responsibility for the content of the manuscript. Separately, this study leverages pre-trained language models (e.g., Sentence-BERT) and large language models (e.g., ChatGPT) as part of the technical methods developed and evaluated in the research, as described in detail in the Methods section.

## I. INTRODUCTION

Form 10-K filings are structured into items to organize firms' disclosures. Important items such as the Management's Discussion and Analysis of Financial Condition and Results of Operations (MD&A; Item 7), Risk Factors (Item 1A), and Legal Proceedings (Item 3) are essential for stakeholders in the capital market. Given their importance, prior accounting research often focuses on the textual content of specific 10-K items to study topics such as information content, market reactions, disclosure quality, and information environment (Loughran and McDonald 2016; Cohen et al. 2020).

Leveraging item-specific text from 10-K filings requires item segmentation, the task of identifying item boundaries and extracting item-specific text. Despite its foundational role in the text analytics pipelines, item segmentation is typically performed using rule-based methods that combine string matching, regular expressions, and HTML parsing (Li 2008; Brown and Tucker 2011; Brown et al. 2020a; Campbell et al. 2014). These methods are brittle and difficult to maintain over time because they must account for substantial variation in item titles, ordering, and document formatting (Loughran and McDonald 2016). Errors introduced at this stage may propagate throughout downstream analytical workflows, degrading model performance and limiting reproducibility.

We argue that two critical gaps remain. First, the literature lacks a standard, large-scale annotated dataset for benchmarking item segmentation performance, hindering systematic evaluation and comparison of alternative approaches. Second, recent advances in deep learning and large language models have not been fully leveraged to address this upstream preprocessing problem. The continued reliance on rule-based systems is particularly vulnerable to regulatory changes and requires continuous manual maintenance. These limitations motivate the development of robust, scalable learning-based item segmentation methods.

In response, we constructed a manually annotated item segmentation dataset comprising 3,737 10-K filings randomly sampled from fiscal years 2001 to 2019. To our knowledge, this dataset is the first large-scale benchmark for training and evaluating item segmentation performance across a wide range of disclosure periods. By enabling systematic comparison of alternative methods under a unified evaluation framework, the dataset addresses a critical infrastructure gap in the literature.

Moreover, building on recent advances in artificial intelligence, particularly large language models (LLMs) and pre-trained language models (PLMs), we develop two novel item segmentation approaches: GPT4ItemSeg and BERT4ItemSeg. GPT4ItemSeg leverages ChatGPT-4o, a LLM with strong semantic understanding, to identify item boundaries in 10-K filings, making it well-suited to handle regulatory changes. To align the model with the item segmentation task, GPT4ItemSeg relies on few-shot prompting, requiring only a small number of partial, manually curated examples to convey the structure and semantics of 10-K items.

This design substantially reduces the need for large labeled training datasets and enables rapid adaptation when new disclosure items are introduced.

At the same time, directly applying LLMs to long regulatory filings presents practical challenges related to context length and hallucination. First, 10-K filings often span tens or even hundreds of thousands of tokens, sometimes exceeding the maximum context window supported by current LLMs. When relevant portions of a filing fall outside the available context, the model may fail to identify correct item boundaries. Second, when LLMs are prompted to generate or reconstruct item-specific text, they may produce fluent but nonidentical content that does not exactly match the original filing, a phenomenon commonly referred to as hallucination. In the context of regulatory disclosures, such deviations are particularly problematic, as even minor discrepancies between the extracted text and the source document can compromise data integrity and downstream empirical analyses.

To address these issues, we introduce the line-ID-based (LIB) prompting mechanism. The LIB approach asks ChatGPT to identify custom-generated line IDs for starting lines of items. A complete item is subsequently generated based on the line IDs and original filings. Our approach addresses the hallucination problem and ensures that the returned item text matches the original filings.

In contrast, BERT4ItemSeg adopts Bidirectional Encoder Representations from Transformers (BERT) (Devlin et al. 2018) to obtain line-by-line embedding and applies Bidirectional Long Short-Term Memory (Bi-LSTM) models (Huang et al. 2015; Hochreiter and Schmidhuber 1997) to learn the label sequences. This PLM, BERT, plays an important role in providing contextualized embedding for each line. Bi-LSTM, on the other hand, is responsible for learning line-level dependencies for accurate item segmentation. The BERT4ItemSeg addresses the limited context window (512 tokens) of BERT by incorporating a hierarchical structure in which the lower-level BERT encoder considers only the context of each line. The upper-level Bi-LSTM, on the other hand, predicts line labels based on the embeddings provided by the lower-level. Since Bi-LSTM has no limits on the sequence length, we effectively overcome the limitation of BERT's short context window.

Our experimental results show that, for items often adopted in previous studies (i.e., Items 1, 1A, 3, 7), the BERT4ItemSeg achieves a macro-F1 of 0.9826, outperforming GPT4ItemSeg (macro F1=0.9552) and baseline approaches such as conditional random field (CRF) (macro F1=0.9788) and rule-based method (0.9048). For other items (excluding Item 15 because of its ambiguous ending position), the BERT4ItemSeg achieves a macro-averaged F-1 of 0.9692 while GPT4ItemSeg achieves a performance level of 0.9385. These results provide high-quality item segmentation that is usable in downstream tasks.

Our study makes multiple contributions with important implications. First, unlike the rule-based approach that relies on manually compiled rules, GPT4ItemSeg and BERT4ItemSeg are based on large language models and pre-trained language models that leverage recent advancements in deep learning and AI to address issues caused by variation in title wording, order, and document formatting, achieving better item segmentation performance.

Second, the proposed GPT4ItemSeg overcomes hallucinations and the limited context window with the novel LIB prompting mechanism. BERT4ItemSeg addresses the limited context window problem by incorporating BERT and Bi-LSTM in a hierarchical structure. While these innovations can also be applied to other long business documents, they originate from the 10-K item segmentation problem, which is fundamental and critical for text analytics in accounting and finance.

Third, GPT4ItemSeg requires fewer than 10 partial examples (i.e., few-shot prompting) to align ChatGPT effectively with the item segmentation task. While the fully trained BERT4ItemSeg performs better than GPT4ItemSeg, the latter is highly effective in scenarios where new SEC regulations introduce additional items. GPT4ItemSeg can be utilized to generate new training examples for BERT4ItemSeg and other machine learning models, enabling efficient operation with smaller models while maintaining flexibility to address concept drift.

Moreover, our findings have critical implications for empirical research in accounting and finance. High-quality and publicly available item segmentation routines directly enhance the reliability and reproducibility of textual analyses, enabling researchers to construct cleaner and consistent datasets to study the links between 10-K disclosures and firm performance, risk assessments, and stock market reactions. This is especially vital in light of the current lack of standardization in text preprocessing pipelines. Practitioners, such as financial analysts, auditors, and institutional investors, can also benefit from these advancements by streamlining the extraction of relevant disclosures, thereby supporting more timely and accurate decision-making.

Finally, the proposed approach offers a generalizable framework for processing other filings. The LIB prompting strategy can be readily adapted to new filing types with minimal manual examples, enabling high-quality segmentation while mitigating the risk of hallucination. In cases where it is economically viable, users can also collect a sufficient number of training instances and fine-tune the BERT4ItemSeg model for the target filing type, thereby facilitating large-scale deployment without incurring ChatGPT API usage costs.

The remainder of this paper is organized as follows. Section II summarizes the related literature. Section III proposes our research questions. Section IV presents the proposed GPT4ItemSeg and BERT4ItemSeg approach. Section V summarizes the experimental results, and Section VI concludes.

# II. LITERATURE REVIEW

Text analytics using 10-K filings is an emerging research field in accounting and finance. Since there are excellent survey papers (Loughran and McDonald 2016; Loughran and McDonald 2020; Das 2014; Kearney and Liu 2014; Li 2010b; Bochkay et al. 2023) on this subject, our discussion focuses on firm disclosures and their theoretical foundations, followed by the item segmentation pipelines and a discussion on LLMs and their applications in accounting.

### Firm Disclosure and 10-K Filings

Information asymmetry between firms and investors has long been recognized as a central issue for an efficient capital market (Healy and Palepu 2001). Public disclosures, such as the annual 10-K filings, aim to reduce information asymmetry by providing information on firms' performance, risks, and governance. Past studies have approached this vital issue from various perspectives, including readability, disclosure similarity, sentiment, forward-looking statements, and risk factors. These studies often leverage either the entire 10-K filing or specific items as the data source for textual analysis. Below, we provide a succinct summary of these research directions.

Readability refers to the cognitive load on readers when comprehending text (Martinc et al. 2021). Impression management theory (Brennan et al. 2009) hypothesizes that managers may obfuscate disclosure information to reduce readability. Arguments based on Agency theory, on the other hand, predict that managers may voluntarily disclose information to reduce information asymmetry (Baginski et al. 2000). Li (2010a) shows that poor performance firms tend to have complicated annual reports, while annual reports with higher readability disclosures are associated with higher earnings persistence. Moreover, the relation between earnings and the Fog Index mainly stems from the MD&A.

Another stream of research focuses on document similarity in disclosures. Hoberg and Phillips (2010) leverage the description of businesses and products in Item 1 to compute product similarity between two firms with cosine similarity. Using text-based measures based on cosine and Jaccard similarity, Cohen et al. (2020) document a trading strategy that shorts firms with large changes in 10-K and 10-Q reports and buys firms with small 10-K and 10-Q report changes, earning up to 188 basis points of abnormal returns per month. Moreover, a trading strategy that focuses on Risk Factors (Item 1A in 10-K) and Legal Proceedings (Item 2.1 in 10-Q) achieves the highest abnormal returns. Other streams of research examine the information content of sentiment (Loughran and McDonald 2011), forward-looking statements (Muslu et al. 2015), and risk factors (Campbell et al. 2014).

It is worth emphasizing that these studies often rely on the ability to segment the 10-K filing into its constituent items accurately. Improved segmentation enables more reliable textual features and enhances reproducibility. As such, methodological advancements in automated item segmentation represent a foundational contribution to the growing literature on text analytics in accounting and finance.

**10-K Item Segmentation**

As discussed in the previous subsection, many studies focus on one or several items in 10-K filings, such as MD&A (Li 2010a; Kothari et al. 2009; Loughran and McDonald 2016), Risk Factors (Campbell et al. 2014; Lyle et al. 2022; Hope et al. 2016), and Business (Hoberg and Phillips 2010). While the quality of item segmentation may greatly affect subsequent analysis, most accounting and finance studies have adopted basic rule-based approaches to segment 10-K items. The EDGAR system provides mechanisms for users to collect a comprehensive set of files. We provide a brief overview of the data preparation pipeline and a summary of the 10-K item segmentation techniques in the literature.

The data preparation pipeline starts by identifying target documents. Our study focuses on annual reports typically associated with document types 10-K and 10-K405 (Loughran and McDonald 2011). The master index (https://www.sec.gov/Archives/edgar/full-index/) lists all available files and provides a convenient venue to construct a comprehensive list of target files. It is straightforward to collect files from the list using modern crawling tools, such as Python's requests library. Raw files downloaded from the EDGAR website are in the Standard Generalized Markup Language (SGML) format. The main 10-K filing is wrapped in a document session, and it can be in Hyper Text Markup Language (HTML) or pure text (i.e., TXT) format. Unwrapping the 10-K filings from the document session provides the main text files for subsequent analysis.

Most studies go through several text pre-processing steps before item segmentation. One of the most common is the removal of HTML Tables (Loughran and McDonald 2011; Li 2008). The rationale is to remove tables that mostly contain numbers; however, this step may also remove text unrelated to tabular data because the <table> tag may be used to format text instead of organizing numeric data (Loughran and McDonald 2014). Several studies also remove paragraphs with more than 50% non-alphabetic characters and replace HTML tags with blanks (Li 2008).

After transforming 10-K filings into a clean format, the next steps are to segment the text and extract the target items. The basic idea is to identify the starting and ending locations of the target items. For example, to identify MD&A (Item 7), one can use keywords such as "Item 7" and "Managements Discussion" to detect the starting locations; keywords such as "Item 7A," "Quantitative and Qualitative Disclosures," or "Item 8" can identify the ending locations (Li 2008; Brown and Tucker 2011; Brown et al. 2020a; Hope et al. 2016). Keyword searching often results in false positive matches. The order of items provides additional clues for selecting more precise locations (Brown and Tucker 2011). An alternative approach operates directly on HTML files and uses visual cues that help human readers recognize boundaries (Campbell et al. 2014).

Keyword-based item segmentation routines typically yield moderate performance. Brown and Tucker (2011) use a regular-expression-based approach and extract 73% of the MD&A items from 10-K filings. Campbell et al. (2014) combine the regular expression approach with the HTML scoring and extract 65% of Items 1A, 7, and 7A from 10-K filings. The precision of the extracted items was 98%.

We believe that extracting items using rules and regular expressions has the following three problems. First, the current HTML-to-text conversion suffers from loss of formatting information, which may negatively influence the quality of item segmentation. Second, complex regular expression rules must be created due to the wide variety of companies' writing styles, typography, and content. Moreover, the title and format of the items change over time to meet the SEC's requirements. Last but not least, the item extraction method using regular expressions cannot process parts of the 10-K reports due to unforeseen formatting variations. Consequently, we hope to develop advanced approaches that leverage large language models and pre-trained language models to perform item segmentation robustly and accurately.

**Applications of Pre-trained Language Models and Large Language Models in Accounting**

The power of AI algorithms gives machines human-like linguistic intelligence (Zhao et al. 2023). At the core of this technology, language modeling (LM) involves estimating the likelihood of word sequences to predict the probability of subsequent tokens. LM initially emerged through statistical language modeling in the 1990s. Subsequent advancements in neural networks and deep learning have driven substantial improvements in LM capabilities. This progresses from Word2Vec, which utilizes a shallow neural network (Mikolov et al. 2013), to BERT, featuring its transformer architecture and self-attention mechanisms (Devlin et al.

2018). Over time, neural language models (NLMs) and PLMs have increasingly demonstrated exceptional performance in addressing NLP tasks. Among these technical advancements, LLMs represent a significant milestone in the field. By incorporating hundreds of billions of parameters and optimizing them on extensive text corpora, LLMs expand the capacity of PLMs and enhance their effectiveness on various downstream tasks (Zhao et al. 2023). While LLMs have performance advantages over PLMs, LLMs either require specialized and expensive hardware or can only be accessed through dedicated APIs. As a result, PLMs continue to play a crucial role in research projects.

The development of domain-specific PLMs such as FinBERT has driven recent advances in sequential labeling for financial text processing. FinBERT (Huang et al. 2023) is a domain-specific adaptation of BERT, pre-trained on financial texts including 10-Ks, 10-Qs, financial analyst reports, and earnings conference call transcripts. This specialization enables it to capture financial terminology and context more effectively, leading to improved performance in sentiment classification and ESG discussion classification. Loukas et al. (2022) developed the FiNER dataset for numerical entity recognition. Their dataset contains more than one million sentences from 10-Ks and 10-Qs, with ground truth identified by iXBRL tags. They have also explored the benefit of continuous pre-training and released SEC-BERT, a BERT model pre-trained on 200K financial filings.

Li et al. (2025) propose a framework to extract accounting data from PDF documents. Their framework converts PDFs to plain text and leverages an LLM for understanding tables of contents and extracting financial variables. Vakilzadeh and Wood (2025) develop a retrieval-augmented generation (RAG)-based AI application to help scholars synthesize accounting research papers more efficiently. Other relevant studies include detecting suspicious transactions (Wang et al. 2025) and key audit matter sentiment analysis (Liu et al. 2022) using deep-learning approaches.

One of the most prominent LLMs is ChatGPT launched by OpenAI in November 2022. Specifically, GPT-3 has 175 billion parameters and was pre-trained on general text data (Yenduri et al. 2024). Compared to GPT-2, which has 1.5 billion parameters, GPT-3 demonstrates a significant increase in capacity achieved by scaling its generative pre-training architecture. Additionally, GPT-3 introduced the concept of in-context learning (ICL), wherein LLMs can be guided to understand new tasks by presenting training examples directly within the prompt (Liu et al. 2021). In March 2023, OpenAI released GPT-4, which extended input capabilities from text to multimodal signals. GPT-4 demonstrates superior performance in addressing complex, human-generated problems and improves safety in responding to provocative or sensitive queries.

ChatGPT's advanced human interaction and problem-solving capabilities have also drawn considerable attention from accounting researchers. Harris (2024) adopts ChatGPT-4 to analyze 10-K filings and develop a metric to measure managers' perception of market competition. The empirical results reveal a positive association between this newly constructed metric and accruals-based earnings manipulation, highlighting the influence of perceptions of market competition on financial reporting practices. Andreou et al. (2021) adopt ChatGPT to analyze disclosure in MD&A, revealing that firms providing more forward-looking disclosures related to R&D activities exhibit a positive association with future idiosyncratic stock price crashes. In addition, the sentiment analysis performance of ChatGPT-4o and ChatGPT-4o mini in the MD&A sections outperforms that of traditional machine learning models and dictionary-based methods (Lehner 2024).

Ethical considerations in AI-generated financial text analysis center on ensuring accuracy, fairness, transparency, and compliance with legal and professional standards (Siau and Wang 2020). Given the high stakes of financial communication, AI outputs must be factually correct and free from hallucinations to prevent misinformation that could distort market behavior.

Users should be informed when content is AI-generated, and human oversight should remain in place for accountability. Privacy and intellectual property rights must be respected, as models may process sensitive or proprietary filings.

# III. RESEARCH QUESTIONS

The value of 10-K items has been widely recognized in previous studies. The lack of advanced 10-K item segmentation approaches and the advancement of LLMs and PLMs provide an opportunity to address this fundamental problem. To systematically investigate this research problem, our Research Question 1 (RQ1) aims to develop a research dataset for training and evaluating 10-K item segmentation models.

Both LLMs and PLMs can effectively address the challenge of 10-K item segmentation. Although both are based on the transformer architecture (Vaswani et al. 2017) and require model pre-training to acquire general linguistic patterns and world knowledge, they differ in several ways. First, LLMs are essentially PLMs with a significantly larger number of parameters. Second, their task-solving paradigms vary. PLMs are primarily designed as transferable task solvers, following a "pre-training and fine-tuning" approach to adapt to specific tasks. LLMs, in contrast, function as general-purpose task solvers. They can handle a wide range of tasks without additional fine-tuning, relying instead on prompt engineering to adapt to the requirements of each task.

Third, LLMs demonstrate a unique capability for few-shot learning through in-context learning. This means they can effectively address tasks with minimal examples provided within the input prompt, a feature not typically supported by smaller PLMs.

While LLMs are promising approaches that lead to state-of-the-art performance on many tasks, they have a few critical shortcomings. LLMs typically have more than 10 billion parameters and are expensive to develop and operate. Second, while LLMs excel in many NLP tasks, they did not perform well in sequence labeling tasks. Recent studies (Ma et al. 2023; Gao et al. 2023) point out that LLMs perform worse or are at best on par with PLMs fine-tuned on appropriate training data. Third, LLMs may hallucinate and generate content that sounds plausible but is, in fact, incorrect. Finally, both PLMs and LLMs are limited in the length of the context window. For example, ChatGPT-4o has a context window of 128 thousand tokens, while PLMs such as BERT only have a context window of 512 tokens. Since a word may map to more than one token, even a 128-thousand-context window may be insufficient.

PLMs and LLMs offer distinct advantages but also have their own weaknesses in item segmentation. We thus plan to investigate their performance and provide concrete guidelines for the most suitable scenarios for using these two approaches. Based on the above discussion, our Research Question 2 (RQ2) is: How do we leverage large language models for 10-K item segmentation? Similarly, our Research Question 3 (RQ3) is: How do we develop a 10-K item segmentation model that leverages pre-trained language models? Finally, Research Question 4 (RQ4) compares and contrasts these approaches and offers specific recommendations for researchers who need to tackle the 10-K item segmentation problem in the future.

## IV. METHODOLOGY

We develop PLM-based and LLM-based approaches to tackle the 10-K item segmentation problem. For the PLM-based approach, we propose BERT4ItemSeg, which leverages BERT and a Bi-LSTM to address the issue of long text. The BERT4ItemSeg approach can handle 10-K reports of arbitrary length and achieve excellent performance. As for the LLM-based approach, we opt to leverage ChatGPT-4o and develop the LIB prompting mechanism to facilitate effective item segmentation and reduce the potential problem caused by hallucination.

Figure 1 presents the overall research framework for our study. Major stages include data collection, manual annotation, PLM-based and LLM-based model development, and performance evaluation. Below, we first discuss the research testbed, which explains the data collection and manual annotation procedure. We then present the BERT4ItemSeg and GPT4ItemSeg approaches, followed by performance evaluation and baseline models.

[Insert Figure 1 Here]

## Research Testbed

We construct the research testbed by randomly sampling 3,737 10-K filings between fiscal years 2001 and 2019. There are 3,400 (91%) filings in HTML format, and the remaining are in pure text format.[1] We preserve the document structures and visual clues when converting the HTML filings to the pure text format.[2] Item headings, the chunk of text in a paragraph, and rows in a table can still be visually recognized after the transformation. We drop a line of text if it contains more than 50% non-alphabetic words.

We recruited 85 master's and undergraduate students from the business school where the first author is employed to annotate documents. All annotators have taken at least one management or accounting-related course. The training process consisted of eight separate three-hour sessions, each involving between five and twelve annotators to allow for closer supervision and more interactive discussion. In each session, the authors first explained the annotation rules in detail, clarifying key concepts, item definitions, and common pitfalls in segmenting and labeling 10-Ks. For these exercises, we prepared 11 annotated filings in advance. Each annotator was assigned six randomly sampled filings from this set and was instructed to compare their own annotations with the reference annotations prepared by the authors. Discrepancies were identified and discussed in real time, with the authors providing explanations, highlighting subtle cases, and addressing edge conditions. Any recurring or ambiguous issues were recorded, and the annotation guidelines were updated accordingly. This iterative process ensured that the annotators progressively refined their skills and internalized the rules. Annotators who successfully completed the training received a cash reward equivalent to 20 US dollars.

---

[1] The dataset is available at https://www.im.ntu.edu.tw/~lu/data/itemseg/itemseg10kdata.7z

[2] We leverage the inscriptis library for the conversion. The library URL is https://github.com/weblyzard/inscriptis. Our systematic exploration shows that other approaches, such as replacing HTML tags with regular expressions or the beautifulsoup library, have problems preserving document structure when converting HTML to pure text format.

Trained annotators were assigned to batches of filings. Each batch contained 50 documents. As a quality-checking mechanism, each batch contained two to four documents that were also assigned to another annotator. The authors checked the annotation quality by computing Cohen's kappa and conducted a detailed review if the value was lower than 0.8. Low-quality annotators were terminated, and their batches were reworked by other annotators. Each annotator was rewarded with cash equivalent to 45 US dollars upon completion of a batch.

The first author randomly sampled and annotated 373 (about 10%) of the reports to ensure the annotation quality. The kappa statistic between the recruited annotators and the first author was 0.92, which is considered reliable. The second author went through the discrepancies and determined the final labels. This subset was designated as the test set (while the remaining reports were used for training and validation) in our subsequent analysis.

Our dataset contains 22 unique items. Item 7 contains an average of 7619.60 words and is the longest item. Item 8, the second longest, contains 7318.01 words, followed by Item 1 (6101.11 words). Table A.1 in Appendix lists the summary statistics for the annotated 10-K reports.

We set up the item label standard by modifying the Beginning-Inside-Outside (BIO) labels (Ratinov and Roth 2009) to include item information. The beginning line of Item 1 is labeled B1, and the subsequent line is I1. Similarly, the beginning and following lines of Item 7 are labeled B7 and I7. Lines not belonging to any of the items are labeled O. Table A.2 in the Appendix lists selected lines of the example 10-K report, together with LineIDs that indicate the line position and the label for each line.

## PLM-based and LLM-based 10-K Item Segmentation

We tackle the 10-K item segmentation problem by leveraging PLMs and LLMs. PLMs and LLMs are the bedrock of current NLP techniques and are widely used in many NLP studies (Minaee et al. 2024). We propose BERT4ItemSeg and GPT4ItemSeg to leverage PLMs and LLMs for 10-K item segmentation. We describe the details in sequence.

### *BERT4ItemSeg*

The BERT4ItemSeg leverages the BERT model (Reimers and Gurevych 2019; Devlin et al. 2018) to encode the contextual information into fixed-length dense vectors. BERT is an encoder-only PLM pre-trained on large corpora using masked language modeling (MLM) and next-sentence prediction tasks (NSP). The MLM pre-training randomly replaces words with a special token ([MASK]). The model then learns to predict the original words based on the surrounding context. In NSP, the model is given two sentences and asked to predict whether the second sentence logically follows the first. Both pre-training tasks help BERT understand context and learn general linguistic patterns.

One severe limitation of BERT is its 512-token context window. Since the average length of a 10-K report is 37,400.92 words, it is impractical to process a full 10-K report using BERT. To address this limitation and accommodate the long document length, we design BERT4ItemSeg with a two-layer structure. The first layer encodes the text line-by-line into a fixed-length embedding vector. The second layer is a Bi-LSTM structure that takes the embedding vectors from each line and predicts item boundaries as labeled using the modified BIO scheme.

The rationale behind BERT4ItemSeg is as follows. First, item boundaries coincide with the beginning of a new line of text. Our preliminary investigation suggests that both HTML- and pure-text-formatted filings follow this pattern. Since our HTML-to-text conversion routine preserves line and paragraph boundaries in the original file, we can leverage this pattern to simplify item segmentation. Second, most lines are shorter than 512 tokens. We can encode individual lines using BERT without hitting the context window length limit. Lines longer than 512 tokens are truncated since the front part of a line should contain enough information for the model to predict item boundaries correctly. Third, since BERT4ItemSeg adopts Bi-LSTM for item boundary prediction, the model can process 10-K filings of arbitrary length, overcoming the limitation imposed by BERT's short context window.

***GPT4ItemSeg***

The rise of LLMs marks the shift from learning task-specific solvers to general-purpose models. A typical approach to utilize LLMs is the in-context learning proposed along with GPT-3 (Brown et al. 2020b). In-context learning formats prompt with a task description, followed by a few examples as demonstrations. The last part of a prompt is the test instance, which, in our case, is the target 10-K report for item segmentation.

Our preliminary investigations using ChatGPT-4o suggest that directly following current in-context learning templates is challenging due to hallucinations and the length of 10-K filings. ChatGPT may deviate from the input 10-K report and generate items different from the original input, even when the prompt explicitly asks the model not to do so. There is no simple workaround for the well-known hallucination problem. Moreover, while the context window of GPT-4o is longer than that of most 10-K reports, adding a few demonstrations at the beginning of the prompt can easily consume a large chunk of the prompt space, squeezing precious space for the final test instance.

We develop the LIB prompting mechanism to address the potential hallucination problem when generating item segmentation results. Moreover, to save prompt space, we compile examples that demonstrate item extraction from partial filings instead of including the full filing. These two critical designs allow us to effectively extract 10-K item boundaries while minimizing the potential downsides of LLMs.

Specifically, the LIB prompting mechanism attaches an integer line ID before each line. The GPT model is asked to output the line ID corresponding to each item's beginning line. The output-parsing program automatically checks whether the output line IDs are integers consistent with the original prompt. Outputs containing inconsistent IDs are rejected and rerun. We construct item boundaries and the item's full text after verifying the line IDs.

In addition to LIB prompting, we format the demonstrations following the same format and provide several examples involving partial 10-K reports. Since text near item boundaries is more informative, we provide text chunks around selected item boundaries along with the corresponding expected outputs. Note that the model can output "NA" for items not in the text.

Table 1 shows the example prompt used in GPT4ItemSeg. We truncate part of the text to save space. The first part of the prompt is the task description: "I am an excellent financial professional. The task is to identify the starting lines of items," followed by an additional explanation regarding possible items in the 10-K. The second part is the few-shot demonstration, providing input and output examples. The last part is the target 10-K formatted using LIB. To reduce the length of longer 10-Ks and allow a full report to fit into the limited prompt space, we truncate the text in a line and keep only the first L words.

[Insert Table 1 Here]

## Performance Evaluation and Baseline Models

We adopt the F1 score as the primary performance indicator for all item segmentation tasks. Item segmentation is similar to a classification task, and the F1 score is a standard performance measure for this type of problem. While we label an item using both the B and I tags, we merge these two tags when computing prediction performance. This approach ensures that all lines within an item are treated equally.

For a target item, the precision is the percentage of correctly classified lines among lines classified as part of the item. The recall of an item is the percentage of lines in the item that have been correctly classified. Since the F1 score is the harmonic mean of precision and recall, we interpret the F1 score of an item as the performance level at which an item can be correctly identified. While our subsequent discussion mainly focuses on the F1 score, we also report average precision, recall, and accuracy. Detailed results, including per-item F1 scores, precision, recall, and accuracy, are provided in the Appendix.

In our experiments, we consider three baselines. The first is a rule-based approach. We develop this approach using a set of regular expressions that capture common item headings and important keywords. The program is implemented as a Python script.

The second baseline is the traditional conditional random field (CRF) model with manually crafted features.[3] Similar to GPT4ItemSeg and BERT4ItemSeg, the unit of analysis is individual lines, and we develop three types of features: semantic, syntactic, and structural.

---

[3] Our implementation is based on the python-crfsuite library. The URL of the library is https://python-crfsuite.readthedocs.io/en/latest/.

The semantic features include unigram and bigram, and the syntactic features include whether the first character is in the upper case and the percentage of upper-case characters in a line. The structural features include word length, the forward position normalized to [0, 1], and the backward position normalized to [0, 1].

## V. EXPERIMENTAL RESULTS

In this section, we first discuss model training and parameter settings. We then present the performance of BERT4ItemSeg, GPT4ItemSeg, and other baseline models, with all prediction metrics computed on a hold-out set. We further examine the role of few-shot examples in GPT4ItemSeg prompts, followed by a dedicated subsection on error analysis to identify common misclassification patterns. In addition, we compare our approaches against alternative public tools to highlight differences in design and applicability. Finally, we summarize the limitations of our study and potential avenues for improvement.

### Model Training and Parameter Setting

GPT4ItemSeg leverages the OpenAI API to access ChatGPT-4o. BERT4ItemSeg is implemented in PyTorch, using a fine-tuned Sentence-BERT model (Reimers and Gurevych 2019) to encode each line of text and feed the resulting embedded vectors into a Bi-LSTM module. We explore the hidden dimensions (64, 128, 256, or 512) and the number of layers (1 or 2) of the Bi-LSTM module. To balance training efficiency and performance under computational constraints, we select a hidden size of 256, two layers, and a batch size of 32. Dropout probability is set to 0.5 after evaluating 0, 0.1, 0.5, and 0.9. Model training is performed using the Adam optimizer with a learning rate of 0.0001. The average length of a line was 25.08 words, and more than 90% of lines had fewer than 100 words. Using the BERT tokenizer, we truncate a line if it is longer than 512 tokens. The early stopping patience parameter is set to 10 epochs. Experiments are conducted on computers with an Intel Core i7-11700 CPU, 128 GB RAM, and an Nvidia RTX 3090 GPU.

**Model Performance**

In this subsection, we present the performance of BERT4ItemSeg and GPT4ItemSeg and compare them with that of the CRF and rule-based approaches. Items 1, 1A, 3, and 7 are often investigated in the literature, and we define them as the "core items" in subsequent discussions. Items with a prevalence lower than 0.7 (see Appendix Table A1 for the statistics) are excluded because low-prevalence items are less important. We have also excluded Item 15 because there are no clear indicators signifying this item's ending. We report a complete set of performance metrics, including F1 score, precision, recall, and accuracy, in Appendix Tables A.3-6.

As listed in Table 2, BERT4ItemSeg achieves a macro-averaged F1 score of 0.9826 for core items and 0.9692 for other items, the highest among all approaches. The CRF, a classic machine learning approach, achieves a macro-averaged F1 score of 0.9788 for core items and 0.9691 for other items. Despite only having access to a few incomplete item segmentation examples in the prompt, GPT4ItemSeg achieves a macro-average F1 score of 0.9552 for core items and 0.9385 for other items. The rule-based approach performs worst, with a macro-averaged F1 score of 0.9048 for core items and 0.9010 for other items.

[Insert Table 2 Here]

As reported in Appendix Table A.6, the rule-based method achieves average precision and recall of 0.9092 and 0.9001, respectively. The precision value is lower than that reported by Campbell et al. (2014), who document an average precision of 98% for Items 1A, 7, and 7A using the HTML scoring method. This discrepancy primarily arises from differences in sample construction. Specifically, Campbell et al. (2014) restrict their analysis to the 65% of HTML filings that can be reliably processed by their rule-based approach, excluding the remaining 35% of filings that fail parsing due to format inconsistencies or structural complexity. In contrast, our evaluation is conducted on a random sample of 10-K filings, without conditioning on file format or structural simplicity, thereby providing a more realistic assessment of item segmentation performance.

We conduct statistical tests on the differences of macro-averaged F1 scores for core items and other items using bootstrap confidence intervals with 5,000 resamples. For the average of core items, the bootstrap results indicate that BERT4ItemSeg significantly outperforms all other approaches under a 95% confidence level. In contrast, for the average of other items, the performance difference between BERT4ItemSeg and CRF is not statistically significant. BERT4ItemSeg significantly outperforms GPT4ItemSeg. Nevertheless, GPT4ItemSeg remains significantly better than the rule-based approach.

For individual items, BERT4ItemSeg performed the best on Items 1, 1A, 2, 6, 7, 7A, 8, 9, and 9A, whereas the CRF performed better on the remaining items. We also note that GPT4ItemSeg exhibits a larger cross-item performance variation compared to other approaches. For example, Item 9A has an F1 score of 0.7903 for GPT4ItemSeg, much lower than other items. Other approaches, such as BERT4ItemSeg and CRF, do not show difficulties processing this item. We suspect this is a result of only having access to the few-shot examples we provided in the prompt and the characteristics of the LLM itself.

Tables 3 and 4 present the F1 score, precision, recall, and accuracy for core items and other items. BERT4ItemSeg outperforms GPT4ItemSeg and other baselines in terms of F1, accuracy, and precision for the core items. CRF achieves the highest recall among all approaches. As for other items, CRF has the highest accuracy and recall, while BERT4ItemSeg performs better in terms of F1 and precision. Consistent with the result of F1 scores for other items, the accuracy difference between BERT4ItemSeg and CRF is not statistically significant, while the pairwise comparison between other approaches is significant. The results indicate that BERT4ItemSeg and CRF achieve roughly the same level of performance, as measured by accuracy. Both approaches, on the other hand, are significantly better than other approaches.

[Insert Table 3 Here]

[Insert Table 4 Here]

To address RQ4, we examine local deployment, usage fees, data protection considerations, and the challenge of concept drift. This discussion concludes with recommendations for researchers and practitioners tackling the item segmentation problem.

Among the evaluated methods, all except GPT4ItemSeg can be deployed locally without requiring access to remote computational resources via APIs. BERT4ItemSeg requires a PC equipped with a consumer-grade GPU, such as an Nvidia GTX 3090 or 4090, to execute the model efficiently. On the other hand, CRF can run on most x86-based PCs with a capable CPU and sufficient memory. The rule-based method has even lower hardware requirements and can operate on modest hardware setups.

Running GPT4ItemSeg incurs token fees for OpenAI API usage, whereas other methods operate entirely locally and therefore incur no additional costs. Another important consideration is data protection. Using GPT4ItemSeg requires submitting prompts and reports to OpenAI's cloud-based API, which means the transmitted content is accessible to the service provider. While 10-K filings are public documents, the specific selection, sequencing, or processing of these filings can reveal nontrivial information, such as analysts' focus on a particular industry, firm, or reporting period. By contrast, running segmentation models locally avoids sending any proprietary context outside the controlled computing environment, thereby offering greater assurance against unintended information leakage.

Concept drift poses another challenge in item segmentation, as the SEC periodically updates its disclosure requirements. For example, the SEC introduced Item 1A in 2005 to mandate risk disclosure and recently announced Item 1C, which requires cybersecurity disclosures for 10-Ks filed for fiscal years ending on or after December 15, 2023. Most models, including BERT4ItemSeg, CRF, and the rule-based method, struggle to adapt to such changes because they cannot recognize new items. Fortunately, GPT4ItemSeg can easily accommodate the situation with minor changes in prompts. To include a new item in the LIB prompting, we need to list the new item as one of the targets in the few-shot demonstrations. If prompt space allows, we can also add a few partial examples highlighting the processing of new items. These changes can be made quickly to allow GPT4ItemSeg to recognize and process new items.

When addressing new items, GPT4ItemSeg is especially valuable for generating additional training examples. These examples can augment the original dataset, enabling a retraining process for BERT4ItemSeg. This approach ensures that researchers and practitioners can maintain access to up-to-date, locally deployable models with robust performance, even as disclosure requirements evolve. By leveraging these recommendations, researchers and practitioners can effectively manage the challenges of item segmentation while ensuring adaptability to changing regulatory landscapes.

Given the evolving nature of SEC disclosure rules, it is essential to establish a model maintenance strategy that balances performance stability with adaptability. Models should be continuously monitored to detect shifts in segmentation performance. While not every SEC update warrants an immediate retraining, significant reporting changes or persistent accuracy degradation should trigger a retraining cycle using updated labeled data. This proactive approach ensures that the segmentation models remain aligned with current reporting standards while minimizing the operational risk of outdated classifications.

Moreover, the proposed approaches can be readily extended to 10-Q reports, which share similar structural patterns and item definitions with 10-K reports. Existing annotation scheme, LIB prompting mechanism, and BERT4ItemSeg model architecture can be applied with minimal modifications. Minor adjustments to the prompt examples in GPT4ItemSeg can accommodate 10-Q–specific items. This adaptability makes the methods suitable for expanding empirical research and automated disclosure analysis beyond 10-Ks.

The approaches proposed in this study have been made publicly available for researchers and practitioners working on 10-K item segmentation at https://github.com/hsinmin/itemseg. BERT4ItemSeg is the preferred choice for users with access to GPU-enabled computers, offering the best combination of performance and deployment flexibility. CRF is a viable alternative for users without access to GPUs, delivering satisfactory results with minor performance trade-offs. We do not recommend the rule-based approach and do not offer this approach in the released open-source tool.

### Few-shot Examples in GPT4ItemSeg Prompts

We conduct extensive experiments to determine the optimal prompt template and the number of few-shot examples. While it is intuitively expected that more examples improve performance, our results show that the gains plateau at around five examples. As shown in Table 5, GPT4ItemSeg's performance increases steadily when the number of examples rises from two to five. However, adding more than five examples does not yield further improvements. Given that the ChatGPT API is billed based on token usage, and that examples must be included in every submission, we selected five few-shot examples as the best balance between performance and cost.

[Insert Table 5 Here]

**Error Analysis**

To better understand the limitations of our approach, we report an error analysis by randomly selecting reports that contain at least one item with an F1 score below 0.8. The patterns observed can be grouped into four main categories: random errors, very short reports, incorrect headings, and unusual item placement. Random errors are misclassifications that show no clear pattern; thus, it is difficult to draw specific recommendations from the model development. While different approaches have random errors at different lines, there are no clear patterns for model improvement. We do note that reports with fewer than 100 lines occasionally underperform. The limited context reduces the amount of information available for the model to identify item boundaries accurately. Moreover, a misclassified line contributes heavily to performance degradation.

Incorrect headings stemmed from mislabeled or unconventional headings. One example is the 2002 10-K filed by Allstates WorldCargo, Inc.[4] In this report, "Controls and Procedures" is labeled as Item 14, whereas the correct heading should be Item 9A. While human annotators identified the inconsistency and labeled the ground truth according to the content rather than the heading, all item segmentation approaches considered in this study misclassified this item as Item 14.

In some filings, items were placed in unconventional locations. For example, the 2006 10-K report filed by Hub International Limited stated in Item 7A that Market Risk is included in

---

[4] https://www.sec.gov/Archives/edgar/data/1072293/000100919102000029/awc10k09-02.txt

Part II of Item 7.[5] However, all item segmentation approaches missed this chunk of Item 7A. The 2007 10-K report filed by Alternative Asset Management Acquisition Corp.[6] stated part of their Item 8 after Item 15. Item segmentation approaches missed this part of Item 8.

These findings highlight that certain "difficult cases" remain challenging for both PLM- and LLM-based approaches, even when overall performance is high. Developing approaches that can robustly handle such irregularities will further improve segmentation accuracy, enhance the completeness of extracted disclosures, and strengthen the reliability of downstream textual analyses.

**Alternative Public Tools**

There are both commercial and open-source alternatives available for downloading and parsing filings from the SEC EDGAR system. Sec-api.io, a commercial tool behind a paywall, provides API services for retrieving 10-K filings and related documents. Edgar Crawler (Loukas et al. 2025) is an open-source project that employs regular expressions to identify and extract specific items from 10-K reports. While both tools offer similar functionalities, there are several key differences between our project (i.e., ItemSeg) and these alternative public tools. We discuss these aspects below.

Our text preprocessing procedure involves three key steps to ensure that the extracted content is clean, consistent, and ready for subsequent analysis. First, the raw HTML files are converted to plain text, followed by an unescaping process to restore characters from their encoded forms. Second, non-basic Latin characters and special symbols are standardized through a translation routine (e.g., "Škoda Auto" to "Skoda Auto,") to reduce inconsistencies that could otherwise affect downstream processing. Finally, rows in numerical tables are systematically removed to eliminate figures that are not the focus of textual analysis and that may introduce noise or distort model performance. Together, these steps produce a text corpus that aligns with the conventions of prior research and facilitates reliable text-based analysis.

[5] https://www.sec.gov/Archives/edgar/data/1133016/000119312507039173/d10k.htm
[6] https://www.sec.gov/Archives/edgar/data/1393816/000089109208001810/e31032_10k.htm

Edgar Crawler partially covers the preprocessing routine, including selected special character translation and the removal of numerical tables. Sec-api.io does not remove rows in numerical tables; instead, it provides a special string to mark the original HTML <table> tag's location. Moreover, characters escaped according to the HTML standard are not restored. Both issues may cause problems in downstream tasks. As an illustrative example, Table 6 provides a side-by-side comparison between ItemSeg and Edgar Crawler for Item 1 of 2023 Apple Inc. 10-K report. ItemSeg converts special and non-basic Latin characters (e.g., ®) to plain text while Edgar Crawler keeps most of these characters. These special characters may introduce noise to downstream tasks.

[Insert Table 6 Here]

Table 7 provides another example of Item 7 output. The output of sec-api.io still keeps unescaped HTML characters (e.g., "Item 7. Management’s Discussion…") while our text preprocessing routine has converted the apostrophe to an appropriate character. It is also clear in this example that sec-api.io added special strings (##TABLE_START and ##TABLE_END) to mark the original positions of the HTML <table> and </table> elements. This practice is less desirable for text analytic tasks in accounting research for two reasons. First, this approach cannot identify tables in the pure-text 10-K reports that are common in the early 2000s. Second, the HTML <table> tag can be used to structure the format instead of showing tables with numerical content. Blindly removing HTML tables may accidentally delete important text content that is essential in downstream tasks.

[Insert Table 7 Here]

The next key difference lies in the underlying technical approaches. Our project offers three learning-based segmentation methods: GPT4ItemSeg, BERT4ItemSeg, and CRF.[7] We do not intend to offer the rule-based method, as our other approaches have demonstrated superior effectiveness. In contrast, Edgar Crawler offers only a rule-based approach for 10-K item segmentation, and sec-api.io does not disclose the methodology underlying its segmentation process. As noted earlier, rule-based approaches are inherently difficult to maintain, particularly when faced with continually evolving disclosure regulations.

We also note that GPT4ItemSeg can accommodate future regulation changes by modifying examples in the few-shot prompts. Once enough item segmentation results are collected, BERT4ItemSeg and CRF can be retrained to accommodate new formats. A similar workflow can be applied to extend GPT4ItemSeg to other filing types. Edgar Crawler, however, needs to engage in system development to accommodate future format changes.

One of our contributions is to provide a manually annotated dataset for 10-K item segmentation and to evaluate segmentation performance systematically. In doing so, we are able to deliver verified technical approaches and make a meaningful contribution to the academic community. However, due to differences in text preprocessing routines, our dataset cannot be directly applied to and evaluated against the other two tools. While our manual inspection suggests that all approaches yield broadly consistent predictions of item boundaries, we caution that this observation is based on a limited number of randomly sampled 10-K reports and may not capture more challenging or atypical cases.

---

[7] Upon acceptance of the manuscript, we plan to release the source code as an open source project that can be installed and used with standard Python library management practices.

Our technical approaches can be applied in different flavors of computer hardware. For basic usage, the CRF can be executed on low-end PCs. BERT4ItemSeg requires a GPU to be executed at a reasonable speed. For users willing to pay token fees, GPT4ItemSeg may be a suitable option. Edgar Crawler uses rule-based approaches, and sec-api.io is invoked through API calls. Both can be executed with low-end PCs. However, we caution that sec-api.io raises similar data protection concerns as GPT4ItemSeg, since both rely on cloud services for processing. Users therefore need to carefully evaluate trade-offs among computational cost, ease of use, scalability, and the potential risks of exposing sensitive data to external providers.

To sum up, while both sec-api.io and Edgar Crawler provide useful alternatives for downloading and parsing 10-K reports, their limitations in preprocessing routines, transparency of their technical approaches, and adaptability to regulatory changes make them less suited for academic research than our proposed system. ItemSeg offers a comprehensive preprocessing pipeline that produces cleaner and more consistent text outputs, multiple learning-based segmentation methods that outperform rule-based approaches, and the flexibility to adapt to evolving disclosure formats. Moreover, our contribution of a manually annotated dataset enables systematic evaluation and supports reproducible research. Together, these features position ItemSeg as a robust and academically grounded solution that not only addresses the shortcomings of existing tools but also advances methodological standards in accounting and finance research. For ease of reference, we provide a side-by-side comparison table in Appendix Table A.7.

**Limitations**

Despite the strong performance of the proposed approaches, several limitations warrant consideration. First, while this dataset spans diverse industries and filing styles, it may not fully capture rare formatting patterns, industry-specific terminology, or unique disclosure structures found in less common filings. Consequently, the reported performance metrics may overstate generalizability when applied to filings outside the sampled period or atypical corporate reporting practices. Additionally, the annotation process remains susceptible to variability in human judgment, especially in borderline cases.

Second, both BERT4ItemSeg and GPT4ItemSeg exhibit limitations inherent to their underlying architectures. BERT4ItemSeg is constrained by its reliance on pre-trained embeddings and supervised learning; it requires additional runs of fine-tuning when SEC reporting requirements change substantially. GPT4ItemSeg, while more adaptable to regulatory changes, relies on a cloud-based API, raising concerns about data privacy and long-term service availability, and incurs token fees to use. Furthermore, GPT4ItemSeg's performance may be sensitive to the future revisions of LLMs.

Finally, the evaluation in this study focused on publicly available 10-K reports and did not account for integration with broader downstream tasks, such as sentiment analysis, topic modeling, or event extraction. While improved segmentation accuracy should enhance the reliability of such tasks, the actual gains in end-to-end analytical pipelines remain to be empirically validated.

# VI. CONCLUSIONS

Items in 10-K filings are essential sources of corporate disclosure information, and accurate segmentation of these items is foundational for robust textual analysis in accounting and finance. Developing automatic extraction routines for key items, such as Items 1A and 7, reduces noise, streamlines research design, and mitigates the risk of misclassification errors that can distort empirical findings. In this study, we introduce two novel approaches: BERT4ItemSeg, which combines BERT embeddings with a Bi-LSTM in a hierarchical architecture, and GPT4ItemSeg, which leverages the LIB prompting mechanism to guide large language model outputs. BERT4ItemSeg achieved a macro-averaged F1 score of 0.9826 for core items and 0.9692 for other items, while GPT4ItemSeg also performed strongly despite relying on only a handful of few-shot examples. Both approaches outperformed the rule-based baseline by a large margin, signifying the benefits of advanced deep learning approaches. To the best of our knowledge, this is the first study to leverage PLMs and LLMs in developing item segmentation approaches that can be applied to analyze all 10-K reports disclosed by US public companies.

A key contribution of this work is the creation of a large, high-quality, and publicly available annotated dataset comprising 3,737 manually segmented 10-K filings from 2001 to 2019. This dataset spans multiple industries, disclosure styles, and reporting formats, making it a valuable benchmark for evaluating segmentation algorithms in accounting and finance research. By making both the dataset and our methodologies available, we provide the research community with the resources needed to replicate our results, compare against standardized baselines, and extend this work in a transparent and reproducible manner. This addresses a long-standing challenge in financial text analysis, where undocumented or ad hoc extraction routines have hindered comparability across studies.

The proposed framework is also highly generalizable to other corporate and regulatory filings. The LIB prompting mechanism in GPT4ItemSeg can be adapted to new document types, such as foreign filers' 40-F forms, 10-Qs, 8-Ks, or proxy statements, with a minimal number of few-shot examples, enabling rapid deployment in response to evolving SEC disclosure requirements and concept drift. Similarly, BERT4ItemSeg can be fine-tuned for alternative filings to produce robust, locally deployable models without recurring API costs. These capabilities make our approaches valuable not only for researchers but also for practitioners, such as financial analysts, auditors, and institutional investors, who seek the timely and accurate extraction of relevant disclosures.

Looking ahead, future research could extend this work by integrating these segmentation methods into end-to-end analytical pipelines, such as sentiment analysis, topic modeling, tabular data interpretation, and numeric extraction, to evaluate the impact of improved segmentation on substantive findings. Additional efforts could explore active learning to reduce annotation costs, hybrid approaches that combine rule-based heuristics with AI models to further enhance accuracy, and hybrid architectures that leverage the robustness of locally deployed PLMs with the flexibility of prompt-driven LLMs. Another promising direction is examining model explainability to better understand how AI models make segmentation decisions and to improve transparency for end users.

FIGURE 1

Graphical Summary of the Research Framework

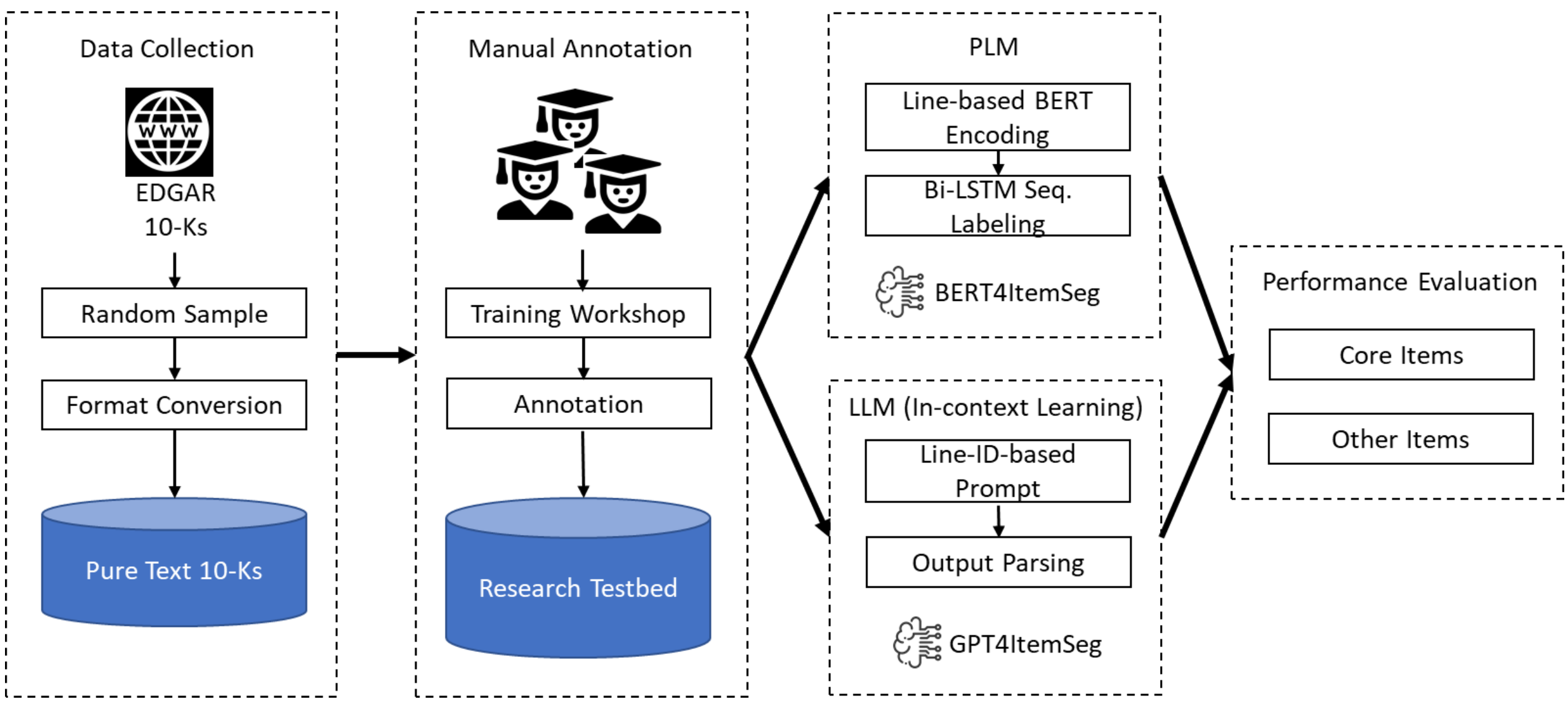

TABLE 1

Example Prompt and Response for 10-K Item Segmentation Using GPT4ItemSeg

**Task Description:**

I am an excellent financial professional. The task is to identify the starting lines of items
in 10-K report.

A 10-K report may contain the following items:
Item 1. Business
Item 1A. Risk Factors
Item 1B. Unresolved Staff Comments
Item 2. Properties
Item 3. Legal Proceedings
Item 4. Mine Safety Disclosures

[…omitted…]

Each item may start with a title, followed by the content. Each line contains a line ID, followed by its content. Extract the line ID of Item 1, Item 1A, Item 2, Item 3, Item 4, Item 5, Item 6, Item 7, Item 7A, Item 8, Item 9, Item 9A, Item 10, Item 11, Item 12, Item 13, Item 14, Item 15. If the item is not available, print NA.

The beginning of a report may contain a table of contents that also lists the item heading but is irrelevant.

**Few Shot Demonstrations:**

Below are some examples.

Example 1:
=====
1 10-K 1 knowles20181231-10xk.htm 10-K
2 UNITED STATES
4 Washington, D.C. 20549
6 FORM 10-K
7
8 (Mark One)
9
10 For the fiscal year ended December 31, 2018 .
11
12 OF 1934
13 For the transition period from to
14
15 Commission File Number: 001-36102

[…Omitted…]

73 Item 9A. Controls and Procedures 93
74 Item 9B. Other Information 94
76 Item 10. Directors, Executive Officers and Corporate Governance 94
77 Item 11. Executive Compensation 95
78 Item 12. Security Ownership of Certain Beneficial Owners and Management and Related Stockholder Matters 95
79 Item 13. Certain Relationships and Related Transactions, and Director Independence 96
80 Item 14. Principal Accountant Fees and Services 96
82 Item 15. Exhibits and Financial Statement Schedules 96
83 Item 16. Form 10-K Summary 100
84
85

86
87 PART I
88 ITEM 1. BUSINESS
89
90 Unless the context otherwise requires, references in this Annual Report on Form 10-K to "Knowles," the "Company," "we," "our,"
91
92 Our Company
93
94 We are a market leader and global provider of advanced micro-acoustic, audio processing, and precision device solutions,
95
96 Our Strategy
=====

Output:
Item 1,87
Item 1A,NA
Item 2,NA
Item 3,NA
Item 4,NA
Item 5,NA
Item 6,NA
Item 7,NA
Item 7A,NA
Item 8,NA
Item 9,NA
Item 9A,NA
Item 10,NA
Item 11,NA
Item 12,NA
Item 13,NA
Item 14,NA
Item 15,NA

Example 2:
=====
[…Omitted…]

## The Task:

Below is a 10-K report.
List the result in a table format. \
The first column is the item ID. The second column is the Line ID. \
Use comma (",") to separate the two columns. Include no additional white space.

=====
0 Table of Contents
1 UNITED STATES
2 SECURITIES AND EXCHANGE COMMISSION
3 x ANNUAL REPORT PURSUANT TO SECTION 13 OR 15(d) OF THE SECURITIES EXCHANGE ACT OF 1934
4 For the fiscal year ended December 31, 2014
5 TRANSITION REPORT PURSUANT TO SECTION 13 OR 15(d) OF THE SECURITIES EXCHANGE ACT OF 1934
6 For the transition period from to
7 Commission File Number: 1-13087
8 BOSTON PROPERTIES, INC.
9 (Exact name of Registrant as specified in its charter)
10 (State or other jurisdiction

11 of incorporation or organization)
12 (I.R.S. Employer
13 Identification Number)
14 Prudential Center, 800 Boylston Street, Suite 1900
15 Boston, Massachusetts
16 (Address of principal executive offices)          (Zip
17 Registrant s telephone number, including area code: (617) 236-3300

[…Omitted…]

3485 /s/ DAVID A. TWARDOCK
3486 David A. Twardock
3487 Director
3488 /s/ MICHAEL E. LABELLE
3489 Michael E. LaBelle
3490 Senior Vice President, Chief Financial Officer and Principal Financial Officer
3491 /s/ LORI W. SILVERSTEIN
3492 Lori W. Silverstein
3493 Vice President, Controller and
3494 Principal Accounting Officer
=====
Output:

**ChatGPT Output:**

```plaintext
Item ID,Line ID
Item 1,67
Item 1A,277
Item 2,558
Item 3,794
Item 4,796
Item 5,800
Item 6,823
Item 7,891
Item 7A,2061
Item 8,2082
Item 9,3126
Item 9A,3128
Item 10,3135
Item 11,3137
Item 12,3139
Item 13,3165
Item 14,3167
Item 15,3171
```

TABLE 2

Comparison of F1 Scores for 10-K Item Segmentation Using GPT4ItemSeg, BERT4ItemSeg, CRF, and a Rule-Based Method.

| **Item** \ **F1** | GPT4ItemSeg | BERT4ItemSeg | CRF | Rule-Based |
|---|---|---|---|---|
| **Core Item** | | | | |
| Item 1 | 0.9672 | **0.9917** | 0.9828 | 0.8966 |
| Item 1A | 0.9388 | **0.9862** | 0.9812 | 0.9118 |
| Item 3 | 0.9503 | 0.9599 | **0.9610** | 0.9054 |
| Item 7 | 0.9644 | **0.9924** | 0.9900 | 0.9053 |
| Core Items Average | 0.9552 | **0.9826** | 0.9788 | 0.9048 |
| **Other Item** | | | | |
| Item 2 | 0.9784 | **0.9880** | 0.9709 | 0.8911 |
| Item 4 | 0.8615 | 0.9015 | **0.9093** | 0.8703 |
| Item 5 | 0.9841 | 0.9912 | **0.9946** | 0.9147 |
| Item 6 | 0.9773 | **0.9883** | 0.9689 | 0.9248 |
| Item 7A | 0.9598 | **0.9707** | 0.9690 | 0.8865 |
| Item 8 | 0.9163 | **0.9468** | 0.9228 | 0.8698 |
| Item 9 | 0.9360 | **0.9801** | 0.9790 | 0.8921 |
| Item 9A | 0.7903 | **0.9623** | 0.9568 | 0.9188 |
| Item 10 | 0.9676 | 0.9895 | **0.9908** | 0.9217 |
| Item 11 | 0.9594 | 0.9886 | **0.9967** | 0.9015 |
| Item 12 | 0.9664 | 0.9926 | **0.9967** | 0.9135 |
| Item 13 | 0.9580 | 0.9869 | **0.9919** | 0.9123 |
| Item 14 | 0.9453 | 0.9136 | **0.9503** | 0.8764 |
| Other Items Average | 0.9385 | **0.9692** | 0.9691 | 0.9010 |

TABLE 3.
Average Performance on Core Items (F1, Precision, Recall, and Accuracy) for GPT4ItemSeg, BERT4ItemSeg, CRF, and Rule-Based.

| | GPT4ItemSeg | BERT4ItemSeg | CRF | Rule-Based |
|---|---|---|---|---|
| F1 | 0.9552 | **0.9826** | 0.9788 | 0.9048 |
| Precision | 0.9567 | **0.9910** | 0.9785 | 0.9092 |
| Recall | 0.9743 | 0.9812 | **0.9846** | 0.9001 |
| Accuracy | 0.9891 | **0.9940** | 0.9938 | 0.9565 |

TABLE 4.
Average Performance on Other Items (F1, Precision, Recall, and Accuracy) for GPT4ItemSeg, BERT4ItemSeg, CRF, and Rule-Based

| | GPT4ItemSeg | BERT4ItemSeg | CRF | Rule-Based |
|---|---|---|---|---|
| F1 | 0.9385 | **0.9692** | 0.9691 | 0.8995 |
| Precision | 0.9349 | **0.9712** | 0.9650 | 0.9032 |
| Recall | 0.9558 | 0.9747 | **0.9807** | 0.8956 |
| Accuracy | 0.9825 | 0.9810 | **0.9838** | 0.9556 |

TABLE 5.
GPT4ItemSeg Performance with Varying Few-Shot Example Counts.

| Few-shot Example Counts | Core Items F1 | All Items F1 | Accuracy |
|---|---|---|---|
| 2 | 0.7256 | 0.5963 | 0.6092 |
| 3 | 0.7902 | 0.6675 | 0.6703 |
| 4 | 0.8856 | 0.7383 | 0.7310 |
| 5 | 0.9552 | 0.9424 | 0.9841 |
| 10 | 0.9573 | 0.9488 | 0.9893 |

Table 6. Comparison of Item 1 Output Between ItemSeg and Edgar Crawler.

Target Report: 2023 10-K filed by Apple Inc.

| **ItemSeg** | **Edgar Crawler** |
| --- | --- |
| Item 1. Business<br><br>Company Background<br><br>The Company designs, manufactures and markets smartphones, personal computers, tablets, wearables and accessories, and sells a variety of related services. The Company's fiscal year is the 52- or 53-week period that ends on the last Saturday of September.<br><br>Products<br><br>iPhone<br><br>iPhone (R) is the Company's line of smartphones based on its iOS operating system. The iPhone line includes iPhone 15 Pro, iPhone 15, iPhone 14, iPhone 13 and iPhone SE (R) .<br><br>[--- truncated ---] | Item 1. Business<br><br>Company Background<br><br>The Company designs, manufactures and markets smartphones, personal computers, tablets, wearables and accessories, and sells a variety of related services. The Company’s fiscal year is the 52- or 53-week period that ends on the last Saturday of September.<br><br>Products<br><br>iPhone<br><br>iPhone® is the Company’s line of smartphones based on its iOS operating system. The iPhone line includes iPhone 15 Pro, iPhone 15, iPhone 14, iPhone 13 and iPhone SE®.<br><br>[---truncated ---] |

Table 7. Comparison of Item 7 Output Between ItemSeg and sec-api.io.
Target report: 2023 10-K Filed by Apple Inc.

| **ItemSeg** | **sec-api.io** |
| --- | --- |
| Item 7. Management's Discussion and Analysis of Financial Condition and Results of Operations<br><br>The following discussion should be read in conjunction with the consolidated financial statements and accompanying notes included in Part II, Item 8 of this Form 10-K. This Item generally discusses 2023 and 2022 items and year-to-year comparisons between 2023 and 2022. Discussions of 2021 items and year-to-year comparisons between 2022 and 2021 are not included, and can be found in "Management's Discussion and Analysis of Financial Condition and Results of Operations" in Part II, Item 7 of the Company's Annual Report on Form 10-K for the fiscal year ended September 24, 2022.<br>[--omitted--]<br>The following table shows net sales by reportable segment for 2023, 2022 and 2021 (dollars in millions):<br><br>Net sales by reportable segment:<br><br>Americas<br>[--truncated--] | Item 7. Management’s Discussion and Analysis of Financial Condition and Results of Operations<br><br>The following discussion should be read in conjunction with the consolidated financial statements and accompanying notes included in Part II, Item 8 of this Form 10-K. This Item generally discusses 2023 and 2022 items and year-to-year comparisons between 2023 and 2022. Discussions of 2021 items and year-to-year comparisons between 2022 and 2021 are not included, and can be found in “Management’s Discussion and Analysis of Financial Condition and Results of Operations” in Part II, Item 7 of the Company’s Annual Report on Form 10-K for the fiscal year ended September 24, 2022.<br>[--omitted--]<br>The following table shows net sales by reportable segment for 2023, 2022 and 2021 (dollars in millions):<br><br>##TABLE_START 2023 Change 2022 Change 2021 Net sales by reportable segment: Americas $ 162,560 (4) % $ 169,658 11 % $ 153,306 Europe 94,294 (1) % 95,118 7 % 89,307 Greater China 72,559 (2) % 74,200 9 % 68,366 Japan 24,257 (7) % 25,977 (9) % 28,482 Rest of Asia Pacific 29,615 1 % 29,375 11 % 26,356 Total net sales $ 383,285 (3) % $ 394,328 8 % $ 365,817 ##TABLE_END<br><br>Americas<br>[--truncated--] |